\documentstyle[12pt]{article}
\rightline{FERMILAB-Pub-92/168-A}
\rightline{June 1992}
\rightline{Submitted to {\it Physical Review Letters}}

\def\la{\mathrel{\mathpalette\fun <}}
\def\ga{\mathrel{\mathpalette\fun >}}
\def\fun#1#2{\lower3.6pt\vbox{\baselineskip0pt\lineskip.9pt
  \ialign{$\mathsurround=0pt#1\hfil##\hfil$\crcr#2\crcr\sim\crcr}}}

\def\mpl{{m_{\rm Pl}}}

\def\VEV#1{\left\langle #1\right\rangle}

\begin{document}
\pagestyle{empty}
\begin{center}

\vspace{.2in}
{\bf COSMIC MICROWAVE BACKGROUND PROBES MODELS OF INFLATION} \\

\vspace{.2in}
Richard L. Davis,$^1$ Hardy M. Hodges,$^2$ George F. Smoot,$^3$ \\
Paul J. Steinhardt$^1$ and Michael S. Turner,$^{4,5}$ \\

$^1${\it Department of Physics, \\
University of Pennsylvania, Philadelphia, PA  19104}\\

$^2${\it Harvard-Smithsonian Center
for Astrophysics, Cambridge, MA 02138} \\

$^3${\it Lawrence Berkeley Laboratory, Space Sciences Laboratory \& \\
Center for Particle Astrophysics,\\
University of California, Berkeley, CA  94720} \\

$^4${\it  Departments of Astronomy \& Astrophysics and of Physics,\\
Enrico Fermi Institute,
The  University of Chicago, Chicago, IL 60637}\\

$^5${\it  NASA/Fermilab Astrophysics Center,  Fermi National Accelerator
Laboratory,  Batavia, IL  60510}\\

\end{center}

\vspace{.1in}
\noindent
{\bf ABSTRACT:}
Inflation creates both scalar (density) and tensor
(gravity wave) metric perturbations.
We find that the tensor mode contribution to the CMB anisotropy
on large-angular scales can only exceed that of the scalar mode
in models where the spectrum of perturbations deviates
significantly from  scale invariance (e.g., extended and power-law
inflation models and extreme versions of chaotic inflation).
If the tensor mode dominates at large-angular scales, then the value
of $\Delta T/T$ predicted on $1^\circ$ is less than
if the scalar mode dominates, and, for cold dark matter
models, $b>1$ can be made consistent with the COBE DMR results.

\vspace{.1in}
\noindent

\newpage
\pagestyle{plain}
\setcounter{page}{1}

The recent COBE DMR \cite{1} measurements of large-angular-scale anisotropy
in the cosmic microwave background (CMB) provide important experimental support
for the hot big bang model.
Perhaps the most striking conclusion to be drawn from the COBE DMR data
is that it is consistent with a scale-invariant
spectrum of primordial density (scalar) perturbations
extending well outside the horizon at the epoch of last scattering.

A scale-invariant spectrum is consistent with inflation,
which predicts perturbations generated by quantum fluctuations \cite{2}, and
also with models that generate perturbations by classical effects,
such as theories with cosmic strings, textures, global monopoles,
and non-topological excitations.   Inflation also produces
a spectrum of gravity waves (tensor metric fluctuations) with
wavelengths extending beyond the horizon, providing a possible means for
distinguishing it from the other scenarios.  Recently it was even speculated
that
the anisotropy detected by the COBE DMR
might be largely due to inflation-produced
tensor rather than scalar perturbations \cite{3}.
In this {\it Letter}, we show that
tensor dominance of the CMB quadrupole anisotropy is indeed possible
for a  class of inflationary models.  We find that the ratio of
tensor to scalar contributions is directly tied to the rate of inflationary
expansion and the ``tilt'' of the spectrum of
density perturbations away from scale invariance.  Models
that permit tensor dominance
include extended inflation, power-law inflation and extreme versions of
chaotic inflation.  While the COBE DMR results alone
cannot distinguish tensor from scalar perturbations,
we show how additional measurements on
small-angular scales may distinguish the two.
We also discuss the implications for large-scale structure.

CMB temperature anisotropies on large-angular scales
($\ga 1^\circ$) are produced by
metric  fluctuations through the Sachs-Wolfe effect \cite{4}.
These temperature fluctuations
can be decomposed into spherical-harmonic amplitudes;
for scale-invariant scalar-mode fluctuations, the
quadrupole is given by \cite{5}
\begin{equation}  \label{scalar}
S \equiv  \VEV{a_2^2}_S \equiv \VEV{\sum_{m=-2}^{m=2}|a_{2m}|^2}
 =  \frac{1}{60 \pi} \frac{H^4}{\dot{\phi}^2}
 = \frac{128\pi^2}{45}\frac{V^3}{{V^\prime}^2 m_{Pl}^6},
\end{equation}
where $H$ is the Hubble parameter, $\phi$ is the
scalar field that rolls during inflation, $V(\phi )$ is
its potential, $m_{Pl} = 1.22 \times 10^{19}$~GeV is
the Planck mass, and the final expression
follows from the slow-roll equation of motion,
$3H\dot{\phi}= -V^\prime$.  The rhs is
to be evaluated $N\sim 60$ e-foldings
before the end of inflation, when fluctuations
on CMB length scales crossed outside the horizon
\cite{6}.  The corresponding formula for
tensor fluctuations is \cite{7}:
\begin{equation}  \label{tensor}
T \equiv \VEV{a_2^2}_T  = 7.74 \frac{V}{\mpl^4},
\end{equation}
The ratio of tensor to scalar quadrupole anisotropies is, therefore,
\begin{equation}  \label{ratio}
\frac{T}{S} \equiv \frac{\VEV{a_2^2}_T }{ \VEV{a_2^2}_S } \approx
0.28
\left(\frac{ V' \mpl}{V}
\right)^2\Bigg\vert_{N \sim 60}.
\end{equation}
Note that the coefficients in Eqs.~(\ref{scalar}, \ref{tensor})
were derived assuming strict scale invariance.  Since we will find below
that models with $T/S \ga 1$
deviate from scale invariance, we have numerically computed the
coefficients in Eqs.~(\ref{scalar}, \ref{tensor})
for ``tilted'' spectra and find that
the numerical coefficient in Eq.~(\ref{ratio}) changes
very little ($\la 10\%$) for the tilts consistent with the COBE DMR
results.

\noindent
\underline{Extended \cite{8} and power-law \cite{9} inflation
models} can be described in terms of a potential of the form,
$V(\phi) = V_0 \; {\exp} (-\beta \phi/\mpl )$,
where $\beta$ is constant or slowly time-dependent.
In extended inflation $\phi$ is related to a field
that is coupled to the scalar curvature
(e.g., a dilaton or Brans-Dicke field), which leads to a modification
of Einstein gravity.  The modified gravity action
can be re-expressed via a Weyl transformation
as the usual Einstein action plus a minimally coupled scalar
field ($\phi$) with an exponential potential.  In the simplest
example of extended inflation \cite{8}, $\beta =
\sqrt{64 \pi/(2 \omega +3)}$, where
$\omega$ is the Brans-Dicke parameter.
For an exponential potential, Eq.~(\ref{ratio}) implies:
\begin{equation}  \label{exp}
{T\over S} \approx 0.28 \beta^2 =  \frac{56}{ 2 \omega +3}
\end{equation}
The ratio $T/S \ga 1$ for
  $\omega \la 26$ ($\beta \ga 1.9$).  Interestingly,
$\omega \la 26$ is almost precisely what is required to avoid
unacceptable inhomogeneities  from  big bubbles
in extended inflation \cite{10}.
(Though $\omega\la 26$ is inconsistent
with solar-system limits for Brans-Dicke theory,
these constraints are evaded by giving the Brans-Dicke field
a mass.)

\noindent
\underline{Chaotic inflation models \cite{11}}
typically invoke a potential of the form, $V(\phi) = \lambda \, \phi^p$,
where $\phi  \gg \mpl$ initially, and rolls to $\phi =0$. The
ratio of tensor to scalar anisotropies can be
expressed in terms of $\phi_N$, the value of the scalar field
$N \sim 60$ e-foldings before the end of inflation.  Using the relation,
\begin{equation}  \label{Nchaos}
N(\phi ) = \int_{t_{\rm end}}^{t_N} H dt =\frac{8 \pi}{\mpl^2}
 \int_{\phi_{\rm end}}^{\phi_N} \frac{V}{V'} d \phi =
 \frac{4\pi}{p}\frac{\phi^2}{\mpl^2} - \frac{p}{12},
\end{equation}
where $\phi_{\rm end}^2=p^2\mpl^2/48\pi$,
we find that \cite{star}:
\begin{equation} \label{chaos}
\frac{T}{S} \approx \frac{p}{17.4}\left[ 1 + \frac{p}{720}\right]^{-1},
\end{equation}
where we have set $N=60$.  For the chaotic-inflation
models usually discussed, $p=2$ and $4$,
the scalar mode dominates:  $T/S=0.11$ and 0.23;
however, for extreme models, $p\ga 18$, the tensor
mode could dominate.

\noindent
\underline{New inflation models \cite{12}}
entail slow-roll from $\phi \approx 0$ to
$\phi = \sigma$ down flat potentials of the
Coleman-Weinberg form,
$V(\phi) =  B \sigma^4/2 + B  \phi^4\;[ {\ln} (\phi^2/\sigma^2)-
\frac{1}{2}]$, where $B\simeq 10^{-15}$ for density perturbations
of an acceptable size.  In new inflation $T/S$ also depends upon $\phi_N$;
paralleling the previous analysis,
\begin{equation} \label{Nnew}
N (\phi ) = \frac{8 \pi}{\mpl^2}
 \int_{\phi_{\rm end}}^{\phi_N} \frac{V}{V'} d \phi \approx
 \frac{\pi}{2|\ln (\phi_N^2 /\sigma^2) |}\frac{\sigma^4}{\phi_N^2\mpl^2};
\end{equation}
\begin{equation}  \label{new}
\frac{T}{S} \approx
 \frac{3.2 \times 10^{-4}}  { |\ln (\phi_N^2/\sigma^2) | }
 \left(\frac{\sigma}{\mpl} \right)^4.
\end{equation}
Scalar dominates tensor for $\sigma  \la 10 \mpl$, and,
naively, it would appear that $T/S$ can be made greater than unity for
$\sigma \ga 10 \mpl$. However, one finds that
$\phi_{60}$ is very close to $\sigma$ for $\sigma \ga 10
\mpl$, violating the implicit assumption, $\phi_{60}
\ll \sigma$.  That is, for $\sigma \gg m_{Pl}$,
$\phi$ rolls down the steeper (harmonic) part of the potential close to
the minimum, so that $V(\phi ) \simeq 4B\sigma^2
(\phi - \sigma )^2$,
just as in chaotic inflation
with $p=2$. In this case,  the tensor mode  does not
dominate ($T/S \simeq 0.11$) \cite{other}.

We will now show that $T/S$ cannot be arbitrarily large
by deriving model-independent relations between $T/S$, the rate of
inflation, and the tilt of the density perturbation spectrum away
from  scale-invariance \cite{hardy}.
The ratio of tensor to scalar perturbations is controlled by
the steepness of the potential, $V^\prime \mpl/V$; cf.~Eq.~(\ref{ratio}).
During inflation, this quantity also determines the
ratio of the kinetic to potential energy of the
scalar field \cite{pre}, $\frac{1}{2}{\dot\phi}^2/V
\simeq (V^\prime\mpl /V)^2/48\pi$, which in turn determines the
effective equation of state ($p=\gamma \rho$) and
the evolution of the cosmic-scale factor ($R\propto t^m$):
$\gamma= [\frac{1}{2}{\dot\phi}^2 -V]/[\frac{1}{2}{\dot\phi}^2
+V]$ and $m= 2/3(1+\gamma )$ (during inflation
$\gamma$ and $m$ can vary).  It is simple to show that
the tensor perturbations are characterized
by a power spectrum $|\delta^T_k|^2\propto k^{n-1}$
and the scalar (density) perturbations
by $|\delta^S_k|^2 \propto k^n$,
where $n=(m-3)/(m-1)$.  In the limit of exponential
inflation, $\frac{1}{2}{\dot\phi}^2/V \rightarrow 0$,
$m\rightarrow \infty$, the tensor and scalar
perturbations are scale invariant ($n=1$) \cite{smoke}.

The relationships between $\frac{1}{2}{\dot\phi}^2/V$
and $m$, $m$ and $n$, together with Eq. (\ref{ratio}),
allow us to express the expansion-rate index $m$ and the power-spectrum
index $n$ (for $N\sim 60$) in terms of $T/S$:
\begin{equation}  \label{mn}
m =  14\left(\frac{S}{T}\right) + 1/3 \simeq 14\left(\frac{S}{T}\right) ;
\qquad  n = 1 - {3 \; (T/S) \over 21- \; (T/S)} \simeq 1 -
\frac{1}{7}\left( \frac{T}{S}\right).
\end{equation}
(We remind the reader that the numerical coefficients
here depend upon that in Eq. (\ref{ratio}), which depends weakly
on the ratio $T/S$ for $n \ga 0.5$.)  If the tensor mode is
to dominate---i.e., $T/S\ga 1$---then
$m$ must be less than about 14 and $n$
must be less than about 0.85.  The converse is also true:
In models where the expansion is exponential
and the spectrum is scale invariant, the ratio of tensor to
scalar is very small.  From the fact that inflation must
be ``superluminal'' ($m > 1$), we can use Eq.~(\ref{mn}) to
derive an approximate {\it upper}
bound, $T/S \la 20$ \cite{caveat}.  However,
the COBE DMR \cite{1} bound on the power-spectrum index $n$,
$n=1.1\pm 0.6$, which implies that $n \ga 0.5$ when
$T/S\ga 1$, leads to the stronger limit, $T/S\la 3$ (and $m\ga 5$).
(Doubtless, there are yet stronger bounds on $n$ based
upon structure formation).

We can  now apply these results for the specific models for which we
found $T/S  \ga 1$, extended and chaotic inflation.  In extended
(or power-law) inflation, the power spectrum is tilted according to
 $n\simeq (2\omega -9)/(2\omega -1)$ and $m= (2 \omega +3)/4$. Using
the COBE DMR limit, $n \ga 0.5$,
we find a plausible range, $26 \ga \omega \ga 9$.
For chaotic inflation, $n \approx 1 - p/120$ and $m \approx 240/p$,
leading to  a somewhat extreme range, $60 \ga p \ga 18$.

Tensor contributions have significant implications for CMB measurements.
First, the COBE DMR results alone
do not distinguish scalar from tensor contributions
to the anisotropy; see Fig.~1.  However, the
COBE DMR results, combined with measurements on smaller-angular scales,
might distinguish the two.  The COBE  DMR measurement implies
$\VEV{a_2^2}=(4.53\pm 2.5)\times 10^{-10}$, where
we should keep in mind that this is a measurement of
 $\VEV{a_2^2}_T+\VEV{a_2^2}_S$.
Going to smaller-angular scales, the scalar contribution
to the CMB anisotropy grows relative to the
tensor, but the net contribution to small-angle measurements is diminished
compared to no tensor mode at all; see Fig.~2.   (We are assuming that no
late re-ionization washes out fluctuations on small-angular scales.)
Hence, comparing large- and small-angle anisotropy measurements
can, in principle, separate the scalar and tensor contributions.
(Another possibility for separating the two
is to measure the polarization of the CMB anisotropy
as the tensor modes lead to a slight polarization \cite{13}.)

The tensor mode can seriously affect the interpretation of CMB
measurements for large-scale structure, regardless of the form
of dark matter.   As an example,  the
best fit cold dark-matter (CDM) model  to the COBE DMR
results assuming $T/S \ll 1$  has  a  bias factor $b\simeq 1$.
(The bias factor $b \equiv 1/\sigma_8$, where
$\sigma_8$ is the {\it rms} mass fluctuation on
the scale $8h^{-1}\,$Mpc.)
If, however, the tensor contribution to the CMB quadrupole
is significant, then the extrapolated density perturbation
amplitude at $8h^{-1}\,$Mpc  is reduced, and
 the best-fit CDM model has $b>1$; see Fig.~2.
Two related effects combine to increase $b$:
the power spectrum
is tilted (less power on small scales for fixed quadrupole
anisotropy), and scalar perturbations only
account for a fraction of the
quadrupole anisotropy.  We find, very roughly,
\begin{equation}  \label{bias}
b \simeq 100^{(1-n)/2}\sqrt{1+T/S} \simeq 10^{(T/S)/7}\sqrt{1+T/S},
\end{equation}
where ``100'' is the ratio of the scale relevant to
the quadrupole anisotropy, $\lambda \sim 1000h^{-1}\,$Mpc,
to the scale $8h^{-1}\,$Mpc.  For $T/S = 0.53, 1.4, 2.5,$
and 3.3, the bias factor $b=1.4, 2.4, 4.6,$ and 7.8
(and $n = 0.92, 0.78, 0.59$ and 0.44).
While these numbers should only be taken as rough estimates, the trend
is clear:  larger $T/S$ permits larger bias.

In sum, if small-angular-scale measurements find $\Delta T/T$
significantly lower than that extrapolated from the COBE DMR quadrupole
(see e.g., \cite{14}), there are now at least two possible
explanations consistent with inflation.
Either re-ionization has washed out the small-angle
fluctuations, or tensor fluctuations contribute significantly
to the COBE DMR observations.
In the latter case, what can CMB studies tell us
about inflation?  Our analysis suggests
a remarkable conclusion---COBE DMR combined with small-angular-scale
measurements can directly relate the key cosmological parameters
that govern large-scale structure, such as the bias factor $b$
in CDM models  and the power-spectrum index $n$,
to the microphysical parameters that control inflation.

\vskip 1.5 cm

\noindent We gratefully acknowledge
discussions with R. Holman and E.W. Kolb.
We thank L. Kofman for bringing Ref.~\cite{13} to our attention
and P. Lubin for sharing some of his data prior to publication.
This research was supported
by the DOE at Penn (DOE-EY-76-C-02-3071),
at Berkeley (DOE-AC-03-76SF0098), and at Chicago and Fermilab,
by the NSF (PHY89-04035), and by the NASA at Fermilab
(NAGW-2381) and at the Harvard-Smithsonian
Center for Astrophysics (NAGW-931).

\newpage
\begin{center}
{\bf FIGURE CAPTIONS}
\end{center}

\noindent
1.  Temperature auto-correlation function
(from the Sachs-Wolfe effect) for  tensor and
scalar modes each normalized to the COBE DMR quadrupole anisotropy
using a scale-invariant spectrum and the COBE DMR window
function \cite{1}. Tensor and scalar modes are distinguishable
at small angles but COBE DMR (data superimposed) is unable
to resolve the difference.
   CDM predictions \cite{15}
for the scalar contribution to $C(0)$ (assuming
$h=0.5$ and $\Omega_b=0.1$) is $C(0) \approx 980$ for
$b=1$ and $C(0) \approx 460$ for $b=1.5$.

\noindent
2. Constraints to the CMB anisotropy from
various experiments and predictions for the South Pole anisotropy
experiment on $1^\circ$ for CDM models
($\Omega =1$, $\Omega_B = 0.1$, $h=0.5$),
using the filter function from \cite{16}:
Open circle, CDM with $b=1$, the best-fit CDM model to
the COBE DMR if $T/S\ll 1$; Open triangle,
CDM with $b=2$, consistent with the COBE DMR only if
$T/S \ga 1$; Closed triangle, upper bound
if COBE DMR were detecting the Sachs-Wolfe effect
from  pure tensor mode ($T/S \gg 1$).


\begin{thebibliography}  {hardy}
\bibitem{6} The comoving scale $\lambda$
crossed outside the horizon when the cosmic-scale factor was
a factor of $e^{-N_\lambda}$ smaller than its value at the
end of inflation:  $N_\lambda \simeq \left(\frac{m}{m-1}\right)
[46 + \ln (\lambda /{\rm Mpc})
+\ln (T_{\rm RH}/10^{14}\,{\rm GeV})]$ ($T_{\rm RH}$ is the
reheat temperature and $R(t) \propto t^m$).
\bibitem{other}  We have examined a variety of other
models, e.g., cosine and polynomial potentials;
the only other example that we found that permits
$T/S \ga 1$ is, $V(\phi )= \lambda (\phi^2-\sigma^2)^2$.
For $\sigma \la 0.8\mpl$, $T/S\ga 1$; since
a necessary condition for sufficient inflation is
$\sigma \ga 0.5\mpl$, this example is a marginal
one at best \cite{pre}.
\bibitem{caveat}  As $T/S$ approaches this
upper bound, $m\rightarrow 1$ and $n\rightarrow -\infty$.
In this limit, our assumptions,
slow roll ($\dot\phi =-V^\prime /3H$)
and the coefficient in Eq. (\ref{ratio}), break down.
\end{thebibliography}
\end{document}